\documentclass[a4paper]{article}
\usepackage{latexsym}
\usepackage{amsthm,amsmath,amssymb}

\usepackage[dvips]{graphicx, color}

\usepackage{color}
\usepackage{bbm}

\newtheorem{thm}{Theorem}

\newtheorem{prop}[thm]{Proposition}

\newtheorem{ex}[]{Example}

\newtheorem{cor}[thm]{Corollary}

\newtheorem{lem}[thm]{Lemma}

\newtheorem{defn}[thm]{Definition}
\newtheorem{rem}[]{Remark}

\newcommand{\cA}{{\mathcal A}}

\newcommand{\cC}{{\mathcal C}}
\newcommand{\cD}{{\mathcal D}}

\newcommand{\cF}{{\mathcal F}}

\newcommand{\cH}{{\mathcal H}}

\newcommand{\cM}{{\mathcal M}}

\newcommand{\cT}{{\mathcal T}}

\newcommand{\A}{{\Bbb A}}

\newcommand{\C}{{\Bbb C}}

\newcommand{\Q}{{\Bbb Q}\hspace{.06em}}
\newcommand{\R}{{\Bbb R}}

\newcommand{\T}{{\Bbb T}}

\newcommand{\Z}{{\Bbb Z}}

\def\ka{\kappa}

\def\={\:=\:}  \def\+{\,+\,}

\def\a{\alpha} \def\b{{\beta}}  \def\ba{\overline\a}

\def\be{\begin{equation}}   \def\ee{\end{equation}}
\def\bes{\begin{equation*}}   \def\ees{\end{equation*}}
\def\ba{\begin{aligned}}   \def\ea{\end{aligned}}
\def\bc{\begin{cases}}   \def\ec{\end{cases}}
\def\bp{\begin{proof}}   \def\ep{\end{proof}}

\def\SL{\mathrm{SL}}

\def\qqan{\qquad\mathrm{and}\qquad}
\def\qan{\quad\mathrm{and}\quad}

\newcommand{\dis}{\displaystyle}

\def\smm{\smallsetminus}

\def\La{\Lambda}
\def\la{\lambda}
\def\Ga{\Gamma}
\def\ga{\gamma}
\def\vol{\mathrm{vol}}
\def\Ker{\mathrm{Ker}}
\def\GL{\mathrm{GL}}
\def\SL{\mathrm{SL}}
\def\Sp{\mathrm{Sp}}

\def\De{\Delta}
\def\lan{\langle}
\def\ran{\rangle}

\def\lan{\langle}
\def\ran{\rangle}

\def\bbm1{\mathbbm 1}

\def\ka{\kappa}
\def\De{\Delta}
\def\whz{\widehat\zeta}

\def\wh{\widehat}
\def\bm{\begin{matrix}}   \def\em{\end{matrix}}
\def\bpm{\begin{pmatrix}}   \def\epm{\end{pmatrix}}
\def\be{\begin{equation}}   \def\ee{\end{equation}}
\def\bes{\begin{equation*}}   \def\ees{\end{equation*}}
\def\bea{\begin{equation}\begin{aligned}}   
\def\eea{\end{aligned}\end{equation}}

\def\whx{\wh\xi}

\def\Lie{\mathrm{Lie}}

\def\SO{\mathrm{SO}}

\begin{document}
\title{\bf\large  Non-Abelian Zeta Function, Fokker-Planck Equation and Projectively 
Flat Connection} 
\author{Lin Weng}  
\date{(March 09, 2019)}
\maketitle
\begin{abstract}
Over the moduli space of rank $n$ semi-stable lattices $\La$ is a universal family 
of tori. Along the fibers, there are natural 
differential operators  and differential equations,  particularly, the heat equations 
and the Fokker-Planck equations in statistical mechanics. In this paper, we 
explain why, by taking averages over the moduli spaces, all these are 
connected with the zeros of rank $n$ non-abelian zeta functions of the field of 
rationals, which are known lie on the central line except a finitely many if $n\geq 2$. 
Certainly, when $n=1$, our current work recovers that of Armitage, which from the 
beginning motivates ours. However,  we reverse the order of the results and the 
hypothesis in their works, i.e. we construct averaged versions of Fokker-Planck 
equations using the above structure of non-abelian zeta zeros. This then leads to an
infinite dimensional Hilbert vector bundle with smooth sections parametrized by non-
abelian zeta zeros. We conjectures that the above structure of Fokker-Planck equation 
comes naturally from an \lq essential projectively flat connection' of the infinite 
dimensional Hilbert bundle and the above family of smooth sections are its \lq essential 
pro-flat sections'. 

\end{abstract}
\tableofcontents

\section{Non-abelian Zeta Functions}
\subsection{Definition}

By definition, a {\it rank $n$ lattice} $\La$ is a discrete subgroup of full rank in an 
$n$-dimensional Euclidean space. It is called {\it semi-stable} if for any sub lattice 
$\La_1$ of $\La$, we have
\be
\vol(\La_1)^n\geq \vol(\La)^{n_1}.
\ee
Here, $n_1$ denotes the rank of $\La_1$ and $\vol(\La)$ and  $\vol(\La_1)$ 
denote the (co)volumes of $\La$ and $\La_1$, respectively.

Denote by $\cM_{\Q,n}$ be the moduli space of semi-stable lattices of rank $n$, 
parametrized the isometric classes of all rank $n$ semi-stable lattices. It is well 
known that  (see \cite{W})
\begin{enumerate}
\item [(1)] There is a natural smooth fibration $\vol: \cM_{\Q,n}\to \R_{>0}$, 
sending the isometric classes $[\La]$ of semi-stable lattices $\La$  to their 
volumes $\vol(\La)$.
\item [(2)] For $t\in \R_{>0}$, denote the fiber of $\vol$ over $t$ by 
$\cM_{\Q,n}[t]$.  Obviously, $\cM_{\Q,n}[t]$ coincides with the moduli spaces of 
semi-stable lattices of rank $n$ and volume $t$. Moreover, as a topological space,  
$\cM_{\Q,n}[t]$ is compact.
\item [(3)] The  natural embeddings 
$\cM_{\Q,n}\hookrightarrow \GL_n(\Z)\backslash \GL_n(\R)$ induces compatible 
embeddings $\cM_{\Q,n}[t]\hookrightarrow \SL_n(\Z)\backslash \SL_n(\R)$. Thus, 
induced from the (Tamagawa) measures on the quotient spaces, we obtain natural 
volume forms $d\mu(\La)$ and $d\mu_t(\La)$ on $\cM_{\Q,n}$ and 
$\cM_{\Q,n}[t]$, respectively. Furthermore, we have 
\be
d\mu(\La)=d\mu_0(\La)\times\frac{dt}{t}.
\ee
\end{enumerate}
With all these, we are ready to introduce the following
\begin{defn}[\cite{W}] The rank $n$ non-abelian zeta function $\wh\zeta_{\Q,n}(s)$ 
of the field $\Q$ of rationals is defined by the integration
\be
\wh\zeta_{\Q,n}(s):=\int_{\cM_{\Q,n}}\left(\theta_\La-1\right)
\cdot\vol(\La)^s d\mu(\La)\qquad\Re(s)>1
\ee
Here, as usual, $\dis{\theta_\La:=\sum_{x\in\La}e^{-\pi|x|_\La^2}}$ denotes the 
theta series of the lattice $\La$.
\end{defn}

Obviously, when $n=1$, $\wh\zeta_{\Q,n}(s)$ coincides with the (completed) 
Riemann zeta function.

\subsection{Riemann Hypothesis}
The non-abelian zeta function $\wh\zeta_{\Q,n}(s)$ satisfies the Riemann 
properties for zeta functions. To explain this, we begin with the structural 
formula for non-abelian zeta function $\whz_{\Q,n}(s)$
\be
s(s-1)\whz_{\Q,n}(s)=s(s-1)\big(I(s)+I(1-s)\big)+\vol(\cM_{\Q,n}[1])
\ee
where
\be\label{eq004}
I(s):=\int_{\cM_{\Q,n}[\leq1]}\left(e^{h^0(\Q,\La)}-1\right)\cdot\vol(\La)^s\,d\mu(\La),
\ee
and $h^i(\Q,\La)\ (i=0,1)$ is the $i$-th arithmetic cohomology of the lattice $\La$ 
characterized by
\be
h^0(\Q,\La)=\log\left(\sum_{x\in\La}e^{-\pi|x|_\La^2}\right).
\ee
Since
\be
\begin{matrix}\cM_{\Q,n}[t]&\simeq& \cM_{\Q,n}[1]\\[0.4em]
\La&\mapsto&t^{1/n}\La\end{matrix}
\ee
we have
\be
\ba
I(s)=&\int_0^1 t^{s}\frac{dt}{t}
\int_{\cM_{\Q,n}[1]}\left(e^{h^0\left(\Q,\frac{1}{t^{1/n}}\La\right)}-1\right)\,d\mu(\La)\\
=&\int_1^\infty t^{-s}\frac{dt}{t}\int_{\cM_{\Q,n}[1]}
\left(e^{h^0\left(\Q,t^{1/n}\La\right)}-1\right)\,d\mu(\La)\\
=&\int_1^\infty t^{-s}\frac{dt}{t}\int_{\cM_{\Q,n}[1]}
\left(\sum_{x\in\La\smm\{0\}}e^{-\pi t^{2/n}|x|_\La^2}\right)\,d\mu(\La)\\
=&\frac{n}{2}\int_1^\infty t^{-\frac{n}{2}s}\frac{dt}{t}\int_{\cM_{\Q,n}[1]}
\left(\sum_{x\in\La\smm\{0\}}e^{-\pi t|x|_\La^2}\right)\,d\mu(\La).
\ea
\ee
Put this back to \eqref{eq004}, we obtain
\be\label{2.6-1}
\ba
&2s(s-1)\whz_{\Q,n}(s)-2\vol(\cM_{\Q,n}[1])\\
=&ns(s-1)\int_1^\infty \left(t^{-\frac{n}{2}s}+t^{-\frac{n}{2}(1-s)}\right)\frac{dt}{t}
\left(\int_{\cM_{\Q,n}[1]}
\Big(\sum_{x\in\La\smm\{0\}}e^{-\pi t|x|_\La^2}\Big)\,d\mu(\La)\right)\\
=&ns(s-1)\int_1^\infty \left(t^{-\frac{n}{2}s}+t^{-\frac{n}{2}(1-s)}\right)\frac{dt}{t}
\left(\int_{\cM_{\Q,n}[1]}\Big(\theta_\La(t)-1\Big)\,d\mu(\La)\right).
\ea
\ee
Hence, at least formally,  we obtain the {\it functional equation} for 
$\whz_{\Q,n}(s)$, i.e.
\be\label{eq0011}
\whz_{\Q,n}(1-s)=\whz_{\Q,n}(s).
\ee
Moreover, by the {\it arithmetic vanishing theorem} on $h^1(\Q,\La)$ and 
arithmetic duality between $h^0(\Q,\La)$ and $h^1(\Q,\La^\vee)$, we can show 
that $I(s)$ is a holomorphic function in $s$. Therefore, the non-abelian zeta 
function $\whz_{\Q,n}(s)$ is well-defined and admits a unique meromorphic 
continuation to the whole $s$-plane, which admits only two singulaties, i.e. two 
simple at $s=0,1$ with the residue at $s=1$ given by  the volume of the compact 
moduli space $\cM_{\Q,n}[1]$. As a by-product, \eqref{eq0011} is justified.
Above all these, much more surprisingly, we have the following result on the 
Riemann hypothesis of $\whz_{\Q,n}(s)$.

\begin{thm}[Theorem 15.4 of \cite{W}]\label{t1} Assume $n\geq 2$. Then all but 
finitely many zeros of $\whz_{\Q,n}(s)$ lie on the  line $\dis{\Re(s)=\frac{1}{2}}$.
\end{thm}

For the basic facts of $\whz_{\Q,n}(s)$, particularly, the proof of this theorem on 
weak Riemann hypothesis for $\whz_{\Q,n}(s)$, please refer to \cite{W}, which  
studies new yet genuine zeta functions for reductive groups over number fields 
systematically.

\section{Heat Equations and Resolvent Formulas}

Recall that a lattice $\La$ consists of a  $\Z$-module $P$ of rank $n$ in the 
$n$-dimensional $\R$-vector space $\R^n$ and an Euclidean metric $H$ on 
$\R^n$. Let $H=(h^{\a\b})$ be the positive definite symmetric matrix associated to 
the rank $n$ lattice $\La$, $x=(x^1,x^2,\ldots,x^n)$ and set
\be
\theta_\La(x,t):=\sum_{\la\in\La}e^{-\pi t|\la|_\La^2}
e^{2\pi i\lan \la,x\ran}
\ee
be the theta function associated to the lattice $\La=(P,H)$. Here, in the equation 
above, we have set 
\be
|(\la_1,\ldots,\la_n)|_\La^2=\sum_{\a,\b=1}^nh^{\a\b}\la_\a\la_\b\qqan
\lan \la,x\ran=\sum_{\a=1}^n\la_\a x^\a
\ee
As usual, we define the {\it dual lattice} $\La^\vee$ of the lattice $\La$ and its 
induced torus $\T_{\La^\vee}^n$ by
\be
\La^\vee:=\big\{x\in\R^n:\lan \la,x\ran\in \Z\ \forall \la\in \La\big\}\qqan 
\T_{\La^\vee}^n:=\R^n/\La^\vee,
\ee
respectively. Then $\theta(x,t)$ may be viewed as a genuine function on the torus
$\T_{\La^\vee}^n$.Since
\be
\frac{\partial}{\partial x^\a}\theta_\La(x,t)
=(2\pi i)\sum_{\la\in\La}e^{-\pi t|\la|_\La^2}e^{2\pi i\lan \la,x\ran} \la_\a,
\ee
we have
\bea
\frac{\partial^2}{\partial x^\b\partial x^\a}\theta_\La(x,t)
=4\pi^2\sum_{\la\in\La}e^{-\pi t|\la|_\La^2}e^{2\pi i\lan \la,x\ran} \la_\a \la_\b
\eea
Therefore,
\bea
\sum_{\a,\b=1}^nh^{\a\b}\frac{\partial^2}{\partial x^\b\partial x^\a}\theta_\La(x,t)
=&4\pi^2\sum_{\la\in\La}e^{-\pi t|\la|_\La^2}e^{2\pi i\lan \la,x\ran}
\sum_{\a,\b=1}^nh^{\a\b}\la_\a\la_\b\\ 
=&4\pi^2\sum_{\la\in\La}e^{-\pi t|\la|_\La^2}e^{2\pi i\lan \la,x\ran}|\la|_\La^2 
\eea
On the other hand, since
\be
\frac{\partial}{\partial t}\theta_\La(x,t)
=-\pi\sum_{\la\in\La}e^{-\pi t|\la|_\La^2}e^{2\pi i\lan \la,x\ran} |\la|_\La^2
\ee
all these then verify the following well-known

\begin{lem}[Heat Equation] On the torus $\T_{\La^\vee}^n$ associated to  the dual 
lattice $\La^\vee$ of the lattice $\La$, the theta function $\theta_\La(x,t)$ of $\La$
is a fundamental solution of the following heat  equation
\be\label{eq.a}
\bc
\dis{\frac{1}{4\pi}\Delta_{\La^\vee,x} \Big(\theta_\La(x,t)\Big)
=\frac{\partial }{\partial t}\Big(\theta_\La(x,t)\Big)},\\[1.07em]
\dis{\hskip 1.80cm \theta_\La(x,0)=\sum_{\la\in\La}e^{2\pi i(\la,x)}}\,.
\ec
\ee
Here  $\Delta_{\La^\vee,x}$ denotes the  Laplacian operator
\be
\Delta_{\La^\vee,x}:=\sum_{\a,\b=1}^nh^{\a\b}\partial_\a\partial_\b
\ee
with $\dis{\partial_\a:=\frac{\partial}{\partial x^\a},\ 
\partial_\b:=\frac{\partial}{\partial x^\b}}.$
\end{lem}

\begin{rem}\normalfont  Even the natural Laplace operator on the tangent bundle 
of $\T_{\La^\vee}^n$ and hence on $\T_{\La^\vee}^n$ is given by
\be
\Delta_{\La^\vee}:=\sum_{i,\a=1}^nh_{\a\b}\partial_\a\partial_\b
\ee
with $(h_{\a\b})$ the inverse matrix of $H=(h^{\a\b})$, for our purpose, we decide 
to use the 'dual'  Laplace operator $\Delta_{\La^\vee,x}$. This  operator 
$\Delta_{\La^\vee,x}$, viewed as the Laplacian on the bundle of differential forms 
of $\T_{\La^\vee}^n$, is also quite natural. Hope that this would not lead to any 
notational confusion. 
\end{rem}

To facilitate our further discussion,  set now $t=:\exp(T),\ x=:\exp(T/2)\,X$,
\be
\Delta_{\La^\vee,X}:=\sum_{\a,\b}h^{\a\b}\frac{\partial^2}{\partial X^\a\partial X^\b},\qquad \Big\lan X,\frac{\partial}{\partial X}\Big\ran
:=\sum_{\a=1}^nX_\a\frac{\partial }{\partial X_\a},
\ee
and
\be
\Theta_\La(X,T):=\theta_\La\left(X\exp(T/2),\exp T\right)
=\sum_{\la\in\La}e^{-\pi e^{T}|\la|_\La^2}e^{2\pi i\lan\la,X\ran e^{T/2}}.
\ee
Then,
\be
\ba
&\Delta_{\La^\vee,x}
=
\sum_{\a,\b}h^{\a\b}\frac{\partial^2}{\partial (X^\a\exp(T/2))\partial (X^\b\exp(T/2))}
=e^{-T}\Delta_{\La^\vee,X},\\[0.40em]
&\frac{1}{4\pi}\Delta_{\La^\vee,X}\Big(\Theta_\La(X,T)\Big)
=-\pi e^T\sum_{\la\in\La}|\la|_\La^2 e^{-\pi e^{T/2}|\la|_\La^2}
e^{2\pi i\lan\la,X\ran e^{T/2}}.
\ea
\ee
On the other hand, since
\bea
&\frac{\partial }{\partial T}\Big(\Theta_\La(X,T)\Big)
=\frac{\partial }{\partial T}
\left(\sum_{\la\in\La}e^{-\pi e^{T}|\la|_\La^2}e^{2\pi i\lan\la,X\ran e^{T/2}}\right)\\
=&\sum_{\la\in\La}e^{\pi e^{T/2}|\la|_\La^2}e^{2\pi i\lan\la,X\ran e^{T/2}}
\left(-\pi e^{T}|\la|_\La^2+\pi i\lan\la,X\ran e^{T/2}\right)\\
=&\frac{1}{4\pi}\Delta_{\La^\vee,X}\Big(\Theta_\La(X,T)\Big)
+\pi i\,e^{T/2}\sum_{\la\in\La}\lan\la,X\ran e^{\pi e^{T/2}|\la|_\La^2}
e^{2\pi i\lan\la,X\ran e^{T/2}},\eea
and
\bea
\Big\lan X,\frac{\partial}{\partial X}\Big\ran
\Big(\Theta_\La(X,T)\Big)
=&\sum_{\b=1}^nX^\b\left(\sum_{\la\in\La}e^{\pi e^{T}|\la|_\La^2}
e^{2\pi i\lan\la,X\ran e^{T/2}}\left(2\pi i\,\la_\b\,e^{T/2}\right)\right)\\
=&2\pi\,i\,e^{T/2}\sum_{\la\in\La}\lan\la,X\ran e^{\pi e^{T/2}|\la|_\La^2}
e^{2\pi i\lan\la,X\ran e^{T/2}} 
\eea
we have proved the following theorem by using the obvious relation  
\be
\Theta_\La(X,0)
=\sum_{\la\in\La}e^{\pi |\la|_\La^2}e^{2\pi i\lan\la,X\ran }=\theta_\La(\la,1).
\ee

\begin{lem} On the dual torus 
$\T_{\La^\vee}\simeq\R^n\big/\left(e^{-T/2}\La^\vee\right)$, we have
\be\label{e1}
\bc
\dis{\left(\frac{\partial }{\partial T}-\Omega_{\La^\vee,X}
\right)}\Big(\Theta_\La(X,T)\Big)=0,\\[0.80em]
\hskip 1.0cm\Theta_\La(X,0)=\theta_\La(x,1).
\ec
\ee
Here, $\Omega_{\La^\vee,X}$ denotes the second order differential operator 
defined by
\be\label{eq0029}
\Omega_{\La^\vee,X}:=\frac{1}{4\pi}\Delta_{\La^\vee,X}
+\frac{1}{2}\Big\lan X,\frac{\partial}{\partial X}\Big\ran.
\ee
\end{lem}

\begin{rm}\normalfont From \eqref{eq0029}, we see that the second differential 
operator $\Omega_{\La^\vee,X}$ is elliptic as well.
\end{rm}

\section{Relation with Non-Abelian Zeta Functions}

Now we are ready to explain why a resolvent version of the discussion above can be naturally connected to  the rank $n$ zeta function. To start with, by \eqref{2.6-1}, in terms of variables $X$ and $T$,  we have
\be\label{e10}
\ba
&2s(s-1)\whz_{\Q,n}(s)-2\vol(\cM_{\Q,n}[1])\\
=&ns(s-1)\int_{\cM_{\Q,n}[1]}
\left(\int_0^\infty \left(e^{\frac{nT}{2}(1-s)}+e^{\frac{nT}{2}s}\right)
\Big(\Theta_\La(0,T)-1\Big)\,dT\right)\,d\mu(\La)
\ea
\ee
Obviously, the integrant for the outer  integration over 
$\cM_{\Q,n}[1]$, namely
\be\int_0^\infty \left(e^{\frac{nT}{2}(1-s)}+e^{\frac{nT}{2}s}\right)
\Big(\Theta_\La(0,T)-1\Big)\,dT,
\ee
involves two Laplace transforms
\be
\int_0^\infty e^{\frac{nT}{2}(1-s)}\Big(\Theta_\La(0,T)-1\Big)\,dT\qan 
\int_0^\infty e^{\frac{nT}{2}s}\Big(\Theta_\La(0,T)-1\Big)\,dT,
\ee
hence there is a contraction semi-group and hence a Feller process induces by 
the heat equation
\be
\Omega_{\La^\vee,X} \Big(\Theta_\La(X,T)-1\Big)
=\frac{\partial }{\partial T} \Big(\Theta_\La(X,T)-1\Big).
\ee
Consequently, we may apply the general resolvent formula  
(see e.g. p.316 of \cite{K}) to obtain 
\be
\int_0^\infty e^{ \kappa T}\Big(\Theta_\La(X,T)-1\Big)dT
=R(\kappa, \Omega_{\La^\vee,X})\Big(\Theta_\La(X,T)-1\Big).
\ee
Here, $R(\kappa,\Omega_{\La^\vee,X})$ denotes the resolvent associated to 
$(-\kappa-\Omega_{\La^\vee,X})^{-1}$. Therefore, we obtain the following
\begin{cor}  For the rank $n$ non-abelian zeta function $\whz_{\Q,n}(s)$, we have
\small{\bea\label{e2}
&\frac{2}{n}\left(\whz_{\Q,n}(s)-\frac{1}{s(s-1)}\vol(\cM_{\Q,n}[1])\right)\\
=&\int_{\cM_{\Q,n}[1]}\!\!\!\left(\left(R\left(-\frac{ns}{2},\Omega_{\La^\vee,X}\right)
\!+\!R\left(\!-\frac{n(1-s)}{2},\Omega_{\La^\vee,X}\right)\right)\!
\Big(\Theta_\La(X,0)-\!1\Big)\right)\Big|_{X=0}\!d\mu(\La).
\eea}
\end{cor}

We end this section by pointing out what are the global structures involved in 
\eqref{e2}. Over the moduli space $\cM_{\Q,n}$ of semi-stable lattices of rank $n$, 
naturally associated is the universal family $\pi:\cT_{\Q,n}^\vee\to\cM_{\Q,n}$, 
characterized by the property that, for each point $[\La]\in  \cM_{\Q,n}$, the fiber 
$\pi^{-1}([\La])$ is simply the torus $\T_{\La^\vee}:=\R^n/\La^\vee$ associated to 
the dual lattice $\La^\vee$ of $\La$. Therefore, on the right hand side of 
\eqref{e2}, namely in
\be\label{2.6-2}
\small{\int_{\cM_{\Q,n}[1]}\left(\left(R\left(-\frac{ns}{2},\Omega_{\La^\vee,X}\right)
+R\left(-
\frac{n(1-s)}{2},\Omega_{\La^\vee,X}\right)\right)\Big(\Theta_\La(X,0)-1\Big)\right)
\Big|_{X=0}
d\mu(\La)},
\ee
the operators $\Omega_{\La^\vee,X}$ are second order elliptic differential 
operators on the fibers $\T_{\La^\vee}$ of $\pi$, the so-called vertical direction, 
while the integration is on the horizontal direction of the base moduli space 
$\cM_{\Q,n}[1]$ of the semi-stable lattices $\La$ of rank $n$. Therefore, the whole 
right hand side is nothing but a horizontal average on the base space of 
$\pi$ for the 'action' of the resolvents on the theta series $\Theta_\La(X,0)$ along 
the vertical fiber direction of $\pi$.

\section{Non-Abelian Zeta Zeros  as Eigenvalues}

In the integration of \eqref{2.6-2}, the integrand is taken along the special section 
$X=0$ in of the fibration $\pi$. But this is artificial: To recover the vertical direction, 
it is enough to assume that $X$ is arbitrary.  As to be expect, for this to work,  a 
price has to be paid with a restriction in continuous parameter $s$, so as to obtain 
certain initial condition to stabilize the solutions of our averaged differential 
equation. It is for this purpose that we have to focus on the zeros of 
$\whz_{\Q,n}(s)$.

By the weak Riemann hypothesis established in Theorem\,\ref{t1}, when 
$n\geq 2$,  all but finitely many zeros of $\whz_{\Q,n}(s)$ lie on the central line 
$\dis{\Re(s)=\frac{1}{2}}$. Let then $\rho=\frac{1}{2}+i \ga\in \frac{1}{2}+i\R$ 
be a zero of $\whz_{\Q,n}(s)$. Then \eqref{e2} becomes
\small{\bea\label{e3}
&\frac{2}{n}\left(\whz_{\Q,n}\left(\frac{1}{2}+i\ga\right)
-\vol(\cM_{\Q,n}[1])\frac{1}{1/4+\ga^2}\right)\\
=&\int_{\cM_{\Q,n}[1]}\!\!\!\left(\!\left(R\left(\!-\frac{n}{4}\!-\!\frac{n}{2}\ga i,
\Omega_{\La^\vee,X}\right)\!+\!R\left(\!-\frac{n}{4}\!+\!\frac{n}{2}\ga i,
\Omega_{\La^\vee,X}\right)\!\right)\!\Big(\Theta_\La(X,0)-1\Big)\!\right)
\!\Big|_{X=0}d\mu(\La)\\
=&\int_{\cM_{\Q,n}[1]}\!\!\!d\mu(\La)\int_0^\infty\!\!\!
\left(e^{(\frac{n}{4}+\frac{n}{2}i\ga)T}\!
+\!e^{(\frac{n}{4}-\frac{n}{2}i\ga)T}\right)\!
\sum_{\la\in\La\smm\{0\}}e^{-\pi e^{T}|\la|_\La^2}\,\!e^{2\pi i\lan\la,X\ran e^{T/2}}
\!\Big|_{X=0}dT.
\eea}
Accordingly, we introduce  a family of functions $\whz_{\Q,n}(X,\frac{1}{2}+i\ga)$ of $X$ parametrized by $\ga$ as follows
\small{\bea\label{e11}
&\frac{2}{n}\left(\whz_{\Q,n}\left(X,\frac{1}{2}+i\ga\right)
-\vol(\cM_{\Q,n}[1])\frac{1}{1/4+\ga^2}\right)\\
:=&\int_{\cM_{\Q,n}[1]}d\mu(\La)\int_0^\infty \left(e^{(\frac{n}{4}+\frac{n}{2}i\ga)T}
+e^{(\frac{n}{4}-\frac{n}{2}i\ga)T}\right)\,\sum_{\la\in\La\smm\{0\}}e^{-\pi e^{T}|\la|_\La^2}
e^{2\pi i\lan\la,X\ran e^{T/2}}dT\\
=&\int_{\cM_{\Q,n}[1]}\left(\left(R\left(-\frac{n}{4}-\frac{n}{2}\ga i,\Omega_{\La^\vee,X}
\right)+R\left(-\frac{n}{4}+\frac{n}{2}\ga i,\Omega_{\La^\vee,X}\right)\right)
\big(\Theta_\La(X,0)-1\big)\right)d\mu(\La)\\
=&4\int_{\cM_{\Q,n}[1]}
\left(R\left(\!-(n\ga)^2,\left(2\Omega_{\La^\vee,X}+\frac{n}{2}\right)^2
\right)\left(\!\left(2\Omega_{\La^\vee,X}+\frac{n}{2}\right)
\big(\Theta_\La(X,0)-1\big)\right)\!\right)d\mu(\La)
\eea
}
Here, in the last step, we have used the following elementary identity
\be
(\Omega-\a)^{-1}+(\Omega-\b)^{-1}
=\big((\Omega-\a)(\Omega-\b)\big)^{-1}\big(2\Omega-(\a+\b)\big).
\ee

To simplify our notations, set
\bea
&\Phi_{\La^\vee}^{(1)}(X,\ga):=\frac{n}{2}\int_0^\infty
\left(e^{(\frac{n}{4}+\frac{n}{2}i\ga)T}+e^{(\frac{n}{4}-\frac{n}{2}i\ga)T}\right)\,
\Big(\Theta_\La(X,T)-1\Big)dT,\\
&\ka_{\La^\vee}^{(1)}(X):=2n\left(2\Omega_{\La^\vee,X}+\frac{n}{2}\right)
\Big(\Theta_{\La}(X,0)-1\Big).
\eea
Since $X$ is a flat section of the fibration $\pi$, the relation \eqref{e11} implies that
\bea\label{e11-2}
&\left(\left(2\Omega_{\La^\vee,X}+\frac{n}{2}\right)^2+(n\ga)^2\right)
\left(\whz_{\Q,n}\left(X,\frac{1}{2}+i\ga\right)
-\frac{\vol(\cM_{\Q,n}[1])}{1/4+\ga^2}\right)\\
=&\int_{\cM_{\Q,n}[1]}\left(\left(\left(2\Omega_{\La^\vee,X}+\frac{n}{2}\right)^2
+(n\ga)^2\right)\Phi_{\La^\vee}^{(1)}(X,\ga)\right)d\mu(\La)\\
=&\int_{\cM_{\Q,n}[1]}\ka_{\La^\vee}^{(1)}(X)d\mu(\La)
\eea
To obtain an initial condition,  note that
\be
\whz_{\Q,n}\left(0,\frac{1}{2}+i\ga\right)=0,
\ee
we hence arrive at the normalizations
\bea\label{e11-2}
\left(\left(2\Omega_{\La^\vee,X}+\frac{n}{2}\right)^2+(n\ga)^2\right)&
\left(\whz_{\Q,n}\Big(X,\frac{1}{2}+i\ga\Big)-\frac{\vol(\cM_{\Q,n}[1])}
{1/4+\ga^2}\Big(1-\exp(-\pi |X|^2)\Big)\right)\\
=&\int_{\cM_{\Q,n}[1]}\left(\left(\left(2\Omega_{\La^\vee,X}+\frac{n}{2}\right)^2
+(n\ga)^2\right)\Phi_{\La^\vee}(X,\ga)\right)d\mu(\La)\\
=&\int_{\cM_{\Q,n}[1]}\ka_{\La^\vee}(X)d\mu(\La).
\eea
where, as to be verified with a long but tedious calculation, the functions 
$\Phi_{\La^\vee}(X,\ga)$ and $\ka_{\La^\vee}(X)$  are given by
\bea
&\Phi_{\La^\vee}(X,\ga)=\Phi_{\La^\vee}^{(1)}(X,\ga)+\frac{1}{1/4+\ga^2}\exp(-\pi |X|^2),\\
&\ka_{\La^\vee}(X)=\ka_{\La^\vee}^{(1)}(X)
+\frac{1}{1/4+\ga^2}\left(\left(2\Omega_{\La^\vee,X}+\frac{n}{2}\right)^2+(n\ga)^2\right)
\exp(-\pi |X|^2)
\eea
In this way, we complete a proof of the following

\begin{thm}\label{th1} Assume that $\dis{frac{1}{2}+i\ga\in\frac{1}{2}+i\R}$ is a 
zero of the non-abelian zeta function $\whz_{\Q,n}(s)$. Then the function 
$\Phi_{\La^\vee}(X,\ga)$ is a solution of the following \lq average' differential 
equation
\be\label{e4}
\int_{\cM_{\Q,n}[1]}\left(\left(\left(2\Omega_{\La^\vee,X}+\frac{n}{2}\right)^2
+(n\ga)^2\right)\Phi_{\La^\vee}(X,\ga)-\ka_{\La^\vee}(X)\right)\,d\mu(\La)=0
\ee
and satisfies the following initial conditions
\begin{enumerate}
\item [(1)] $\dis{\int_{\cM_{\Q,n}[1]}\Phi_{\La^\vee}(0,\ga)\,d\mu(\La)=0}$,
\item [(2)] $\dis{\int_{\cM_{\Q,n}[1]}\frac{\partial}{\partial X_i}
\Phi_{\La^\vee}(0,\ga)\,d\mu(\La)
=\int_{\cM_{\Q,n}[1]}\frac{\partial^3}{\partial X_i^3}\Phi_{\La^\vee}(0,\ga)\,d\mu(\La)=0}$;
\item [(3)] $\dis{\lim_{|X|\to\infty}\left|\int_{\cM_{\Q,n}[1]}\frac{\partial^j}{\partial X_i^j}
\Phi_{\La^\vee}(X,\ga)\,d\mu(\La)\right|=O\left(\frac{1}{|X|}\right)}$ for $0\leq j\leq 4.$
\end{enumerate}
\end{thm}

\bp
Indeed, \eqref{e4} is a direct consequence of  \eqref{e3} and  \eqref{e11-2} as 
mentioned above. Since $\whz_{\Q,n}(1/2+i\ga)=0$, resp. 
$\Phi_{\La^\vee}(X,\ga)=\Phi_{\La^\vee}(-X,\ga)$, we have (1) and (2). In the same 
line, since the involvement of $X$ in $\Phi_{\La^\vee}(X,\ga)$ is rather simple, we 
have (3) as well by direct calculations.
\ep

We may rewrite the integrant in \eqref{e4} more uniformly in the language of 
perturbations. To explain this, we first give some  functional analysis preparations.

Suppose there would be a family of Hilbert spaces $\cH_{\La,0}$ of functions 
defined on the fibers  $\T_{e^{-{T}/{2}}\La^\vee}$ of 
$\pi:\cT_{\Q,n}\to\cM_{\Q,n}[1]$ with associated inner products $(\cdot,\cdot)$, 
containing both $\Phi_{\La^\vee}(X,\ga)$ and $\ka_{\La^\vee}(X)$ for all the 
$\ga$'s. Denote the real and the imaginary part of $\Phi_{\La^\vee}(X,\ga)$ by 
$\a_{\La,\ga}(X)$ and $\b_{\La,\ga}(X)$. Then
\be
\Phi_{\La^\vee}(X,\ga)=\a_{\La,\ga}(X)+i\b_{\La,\ga}(X)\qqan 
\Phi_{\La^\vee}(X,\bar\ga)=\a_{\La,\ga}(X)-i\b_{\La,\ga}(X).
\ee
In the spaces $\cH_{\La,0}$, chose a family of function $\varphi_\La(X)$ such that
\be
<\b_{\La,\ga}(X),\varphi_\La(X)>=0\qqan <\a_{\La,\ga}(X),\varphi_\La(X)>\not=0.
\ee
We then could  define a family of projection operators 
\be
\begin{matrix}P_{\La,\ga}:& \cH_{\La,0}&\to &<\ka_{\La^\vee}(X)>,\\[0.5em]
&f&\mapsto&-\dis{\frac{<f,\varphi_\La(X)>}{<\a_{\La,\ga}(X),\varphi_\La(X)>}}
\cdot \ka_{\La^\vee}(X)\end{matrix}
\ee
Certainly, $P_{\La,\ga}$ depends on $\ga$ and 
\be
P_{\La,\bar\ga}=P_{\La,\ga}.
\ee
\begin{prop} Assume $\dis{\frac{1}{2}+\ga}$ is a zero of \,$\whz_{\Q,n}(s)$ on the 
central line. Then the average on $\cM_{\Q,n}[1]$ of the associated micro 
perturbations of \eqref{e4} becomes stable in the sense that
\be
\int_{\cM_{\Q,n}[1]}
\left(\left(\left(2\Omega_{\La^\vee,X}+\frac{n}{2}\right)^2+P_{\La,\ga}
+(n\ga)^2\right)\Phi_{\La^\vee}(X,\ga)\right)\,d\mu(\La)=0.
\ee
\end{prop}

When, $n\geq 2$,  all but finitely many $\ga$ are real. Hence, by taking an 
average over $\cM_{\Q,n}[1]$, the number $-(n\ga)^2$ is a sort of common real 
eigenvalue of the perturbed operator 
$\dis{\left(2\Omega_{\La^\vee,X}+\frac{n}{2}\right)^2+P_{\La,\ga}}$ on a fixed 
family of Hilbert spaces $\cH_{\La,0}$, or better, a bundle of Hilbert spaces of 
$\pi$.

\section{Fokker-Planck Equations and Beyond}
\subsection{Fokker-Planck Equations}
Consider the basic differential equation stated in Theorem\,\ref{th1}
\be\label{e20}
\int_{\cM_{\Q,n}[1]}\left(\left(\left(2\Omega_{\La^\vee,X}+\frac{n}{2}\right)^2
+(n\ga)^2\right)\Phi_{\La^\vee}(X,\ga)-\ka_{\La^\vee}(X)\right)\,d\mu(\La)=0.
\ee
Since all the functions involved define tempered distributions on the fibers 
$\T_{\La^\vee}$
of the universal families $\pi: \cT_{\Q,n}^\vee\to \cM_{\Q,n}$ at $[\La]$, we may 
take fiberwise  Fourier transforms (for details,  see e.g.  \S\ref{aB} in the the 
appendices). Namely, along each fiber $\T_{\La^\vee}$,
\be
\wh f(Y):=\cF(f(X))
:=\int_{-\infty}^\infty\ldots\int_{-\infty}^\infty e^{-2\pi i \lan Y,X\ran }f(X)\,dX
\ee
where $dX$ is induced from a smooth family of volume forms on $\T_\La$, 
compatible with the fibration structure of the universal family 
$\pi:\cT\to \cM_{\Q,n}$ on $\cM_{\Q,n}$.  Hence the following formulas hold 
tautologically  
\be
\wh{D^af}=(2\pi i x)^a \wh f\qqan \wh{(-2\pi i x)^a f}=D^a\wh f 
\ee
where, for $a=(a_1,\ldots a_n)$,  $\dis{|a|:=\sum_{\a=1}^na_i}$ and 
\be
x^a:=\prod_{\a=1}^nx_i^{a_i}\qqan \frac{\partial^{|a|}}{\partial x^a}
:=\frac{\partial^{|a|}}{\partial x_1^{a_1}\ldots\partial x_n^{a_n}}.
\ee
Consequently, we have
\be
\big(x^\a\partial_\a  f\big)\wh\,=\frac{1}{-2\pi i}\partial_\a\big(\partial_\a  f\big)
\wh\,=\frac{2\pi i}{-2\pi i}\partial_\a (x^\a\wh f)=-\partial_\a (x^\a\wh f).
\ee
Therefore, \lq dual' to \eqref{e20}, with a direct but tedious calculation, we 
conclude that  
\small{\bea\label{e6}
\int_{\cM_{\Q,n}[1]}&\left(\sum_{\a,\b=1}^n\frac{\partial}{\partial Y_\a}
\Big(Y_a\frac{\partial}{\partial Y_\b}\Big( Y_\b \wh\Phi_{\La^\vee}(Y,\ga)\Big)\Big)
-n \sum_{\a=1}^n\frac{\partial}{\partial Y_\a} \left(Y_\a\wh\Phi_{\La^\vee}(Y,\ga)\right)
\right.\\
+2\pi& \Big(\sum_{\a,\b,\tau=1}^nh^{\a\b}Y_\a Y_\b\frac{\partial}{\partial Y_\tau}
\left(Y_\tau \wh\Phi_{\La^\vee}(Y,\ga)\right)
+\sum_{\a,\b,\tau=1}^nh^{\a\b}\frac{\partial}{\partial Y_\tau}
(Y_\a Y_\b Y_\tau \wh\Phi_{\La^\vee}(Y,\ga))\Big)\\
&+4\pi^2\Big(\sum_{\a,\b=1}^nh^{\a\b}Y_\a Y_\b\Big)^2 \wh\Phi_{\La^\vee}(Y,\ga)
-2\pi n  \sum_{\a,\b=1}^nh^{\a\b}Y_\a Y_\b\wh\Phi_{\La^\vee}(Y,\ga)\\
&\hskip 4.30cm \left.+n^2\left(\frac{1}{4}+\ga^2\right) \wh\Phi_{\La^\vee}(Y,\ga)-
\wh{\ka_{\La^\vee}(Y)}
\right)d\mu=0
\eea}

To continue, we next give the following elementary lemma, which can be deduced 
from a long but tedious elementary calculation, using the following elementary 
relation repeatly
\be
Y_a\frac{\partial}{\partial Y_\a}(A)=\frac{\partial}{\partial Y_\a}(Y_\a A)-A
\ee
where $A$ is a $C^1$ function in $Y$.
\begin{lem} The following identities hold.
\begingroup
    \fontsize{8pt}{8pt}\selectfont
    {\bea&\sum_{\a,\b=1}^n\frac{\partial}{\partial Y_\a}
\Big(Y_a\frac{\partial}{\partial Y_\b}\Big( Y_\b\, \wh\Phi_{\La^\vee}(Y,\ga)\Big)\Big)\\
=&n^2\wh\Phi_{\La^\vee}(Y,\ga)+(2n+1)
\sum_{\a=1}^nY_\a\frac{\partial}{\partial Y_\a}\Big(  \wh\Phi_{\La^\vee}(Y,\ga)\Big)
+\sum_{\a,\b=1}^nY_\a Y_\b\frac{\partial^2}{\partial Y_\a\partial Y_\b}
\Big(  \wh\Phi_{\La^\vee}(Y,\ga)\Big),\\
&n\sum_{\a=1}^n\frac{\partial}{\partial Y_\a} \left(Y_\a\wh\Phi_{\La^\vee}(Y,\ga)
\right)
=n^2\wh\Phi_{\La^\vee}(Y,\ga)+n\sum_{\a=1}^nY_\a\frac{\partial}{\partial Y_\a} 
\left(\wh\Phi_{\La^\vee}(Y,\ga)\right),\\
&\sum_{\a,\b,\tau=1}^nh^{\a\b}Y_\a Y_\b\frac{\partial}{\partial Y_\tau}
\left(Y_\tau \wh\Phi_{\La^\vee}(Y,\ga)\right)
=\sum_{\a,\b,\tau=1}^nh^{\a\b}Y_\a Y_\b\left(1+Y_\tau\frac{\partial}{\partial Y_\tau}
\right)
\left( \wh\Phi_{\La^\vee}(Y,\ga)\right),\\
&\sum_{\a,\b,\tau=1}^nh^{\a\b}\frac{\partial}{\partial Y_\tau}
(Y_\a Y_\b Y_\tau \wh\Phi_{\La^\vee}(Y,\ga)\Big)
=-\sum_{\a=1}^nh^{\a\a}Y_\a^2\Big(\wh\Phi_{\La^\vee}(Y,\ga))\Big)\\
&\qquad+(n+3)
\sum_{\a,\b=1}^n h^{\a\b}Y_\a Y_\b\Big( \wh\Phi_{\La^\vee}(Y,\ga)\Big)
+\sum_{\a,\b,\tau=1}^n h^{\a\b}Y_\a Y_\b Y_\tau\frac{\partial}{\partial Y_\tau}
\Big( \wh\Phi_{\La^\vee}(Y,\ga)\Big).
\eea}
\endgroup
Similarly, in terms of backward expression, we have
\begingroup
    \fontsize{8pt}{8pt}\selectfont
    {\bea&\sum_{\a,\b=1}^n\frac{\partial}{\partial Y_\a}
\Big(Y_a\frac{\partial}{\partial Y_\b}\Big( Y_\b\, \wh\Phi_{\La^\vee}(Y,\ga)\Big)\Big)
=\sum_{\a,\b=1}^n\frac{\partial^2}{\partial Y_\a\partial Y_\b}\Big( Y_\a Y_\b 
\wh\Phi_{\La^\vee}(Y,\ga)\Big)\\
&\hskip 6.0cm
-\sum_{\a=1}^n\frac{\partial}{\partial Y_\a}\Big( Y_\a\wh\Phi_{\La^\vee}(Y,\ga) \Big),\\
&\sum_{\a,\b,\tau=1}^nh^{\a\b}Y_\a Y_\b\frac{\partial}{\partial Y_\tau}
\left(Y_\tau \wh\Phi_{\La^\vee}(Y,\ga)\right)
=\sum_{\a,\b,\tau=1}^nh^{\a\b}\frac{\partial}{\partial Y_\tau}
\left(Y_\a Y_\b Y_\tau \wh\Phi_{\La^\vee}(Y,\ga)\right)\\
&\hskip 1.8cm-2\sum_{\a,\b=1}^nh^{\a\b}\left(Y_\a Y_\b\wh\Phi_{\La^\vee}(Y,\ga)
\right)
+\sum_{\a=1}^nh^{\a\a}(Y_\a^2-Y_a)\wh\Phi_{\La^\vee}(Y,\ga).\\
\eea}
\endgroup
\end{lem}

By applying this lemma, from \eqref{e6}, we finally arrive at the following

\begin{thm} Assume that $\frac{1}{2}+\ga$ is a zero of non-abelian zeta function
$\whz_{\Q,n}(s)$ on the central line. Then, we have the following equivalent form 
of the average of Fokker-Planck equations over $\cM_{\Q,n}[1]$.

\begingroup
    \fontsize{8pt}{8pt}\selectfont\small{\bea\label{e6}
&[(1)]\quad\int_{\cM_{\Q,n}[1]}\left(\sum_{\a,\b=1}^n\frac{\partial}{\partial Y_\a}
Y_a\frac{\partial}{\partial Y_\b} Y_\b 
-n \sum_{\a=1}^n\frac{\partial}{\partial Y_\a} Y_\a
\right.\\
&\hskip 2.50cm
+2\pi \Big(\sum_{\a,\b,\tau=1}^nh^{\a\b}Y_\a Y_\b\frac{\partial}{\partial Y_\tau}
Y_\tau 
+\sum_{\a,\b,\tau=1}^nh^{\a\b}\frac{\partial}{\partial Y_\tau}
Y_\a Y_\b Y_\tau \Big)\\
&\left.+4\pi^2\Big(\sum_{\a,\b=1}^nh^{\a\b}Y_\a Y_\b\Big)^2 
-2\pi n  \sum_{\a,\b=1}^nh^{\a\b}Y_\a Y_\b+n^2\left(\frac{1}{4}+\ga^2\right)\right)
\wh\Phi_{\La^\vee}(Y,\ga)\\
&\hskip 7.90cm \left. -\wh{\ka_{\La^\vee}(Y)}
\right)d\mu=0
\eea}

\small{\bea\label{e6-2}
&[(2)]\quad\int_{\cM_{\Q,n}[1]}\!\!\left(\!\!\left(\sum_{\a,\b=1}^nY_\a Y_\b
\frac{\partial^2}{\partial Y_\a\partial Y_\b}+\Big(4\pi\sum_{\a,\b=1}^nh^{\a\b}Y_\a Y_\b 
+ (n+1)\Big)\sum_{\a=1}^nY_\a\frac{\partial}{\partial Y_\a}\right.\right.\\
& \left.+\Big(\Big(2\pi\sum_{\a,\b=1}^nh^{\a\b}Y_\a Y_\b+1\Big)^2
-\Big(2\pi\sum_{\a=1}^nh^{\a\a}Y_\a^2+1\Big)+n^2\left(\frac{1}{4}+\ga^2\right)\right)
\wh\Phi_{\La^\vee}(Y,\ga)\\
&\hskip 9.0cm
\left. -\wh{\ka_{\La^\vee}(Y)}
\right)d\mu=0\\
\eea}
~\\[-3em]
\small{\bea\label{e6-3}
&[(3)]\quad\int_{\cM_{\Q,n}[1]}\left(\!\!\left(\!\!\left(\sum_{\a,\b=1}^n
\frac{\partial^2}{\partial Y_\a\partial Y_\b} Y_\a Y_\b 
-(n+1) \sum_{\a=1}^n\frac{\partial}{\partial Y_\a} Y_a
\Big(1+4\pi \sum_{\b,\tau=1}^nh^{\b\tau}Y_\b Y_\tau\Big)\!\!\right)\right.\right.\\
&\left.+2\pi\!\!\left(\!2\pi\Big(\sum_{\a,\b=1}^nh^{\a\b}Y_\a Y_\b\Big)^2\!\!-\!(n+2)\!
\sum_{\a,\b=1}^nh^{\a\b}Y_\a Y_\b
\!+\sum_{\a=1}^nh^{\a\a}(Y_\a^2-Y_a)\Big)\right)\!\!\right)\wh\Phi_{\La^\vee}(Y,\ga)
\\
&\hskip 5.50cm \left.\left.+n^2\left(\frac{1}{4}+\ga^2\right) \wh\Phi_{\La^\vee}(Y,
\ga)-
\wh{\ka_{\La^\vee}(Y)}\right)\!\!\right)d\mu=0
\eea}\endgroup
\end{thm}

The equations \eqref{e6}, \eqref{e6-2}, \eqref{e6-3} are the variations of the 
equation
\bea
\int_{\cM_{\Q,n}[1]}&\left(\left(\sum_{\a=1}^n\frac{\partial}{\partial Y_\a}
Y_\a\right)^2
\Big(B_2(Y)\cdot \wh\Phi_{\La^\vee}(Y,\ga)\Big)
+\left(\sum_{\a=1}^n\frac{\partial}{\partial Y_\a}Y_\a\right)
\Big(B_1(Y)\wh\Phi_{\La^\vee}(Y,\ga)\Big)\right.\\
&\hskip 3.70cm \left.+B_0(Y)\wh\Phi_{\La^\vee}(Y,\ga)+\ga^2\, \wh\Phi_{\La^\vee}
(Y,\ga)-\wh\ka(Y)\right)
d\mu=0
\eea
for suitable functions $B_0(Y), B_1(Y)$ and $B_2(Y)$ on the fibers of the fibration 
$\pi:\cT_{\Q,n}\to \cD_1$ defined by the follows: for each $[\La]\in\cM_{\Q,n}[1]$, the 
fiber of 
$\pi$ over $[\La]$  is the torus $\T_\La=\R^n/\La$. This may be viewed as an 
average on $\cM_{\Q,n}[1]$ form of  the (backward and/or forward) {\it Fokker-Planck 
equation},  in dimension $n$.

\subsection{Essential Projectively Flat Connection and Pro-Flat Sections}

Consider the functions $\big\{\wh{\Phi_{\La^\vee}(Y,\ga)}\big\}_\ga$ parametrized by by 
the zeros of the rank $n$-non-abelian zeta function $\whz_{\Q,n}(s)$ on the central 
line. Smoothly depending on $Y$, we may view 
$\big\{\wh{\Phi_{\La^\vee}(Y,\ga)}\big\}_\ga$ as an  infinite family of smooth sections
of the infinite dimensional Hilbert vector bundle $\cH_{\whz,n}$ on the universal family 
$\cT\to \cM_{\Q,n}[1]$. Apparently quite complicated, we believe that the Fokker-Planck 
equation reveals some natural geometric differential structure for this infinite 
dimensional Hilbert vector bundle $\cH_{\zeta,n}$. To be more explicit, there should 
exists a Hilbert-metric-compatible projectively flat connection (modulo the integration 
over the base moduli space $\cM_{\Q,n}[1]$) such that 
$\big\{\wh{\Phi_{\La^\vee}(Y,\ga)}\big\}_\ga$ are a base of the projectively flat 
connections again  modulo the integration over the base moduli space $\cM_{\Q,n}[1]$.  
If exists, we call such a connection an essential projectively flat connection and the 
above family a family of essential pro-flat section.  This is the source of positivity
which guarantees that all zeta zeros are on the  central line.

\section{Examples}
\subsection{Rank One}

In the case, when $n=1$, the above work is nothing but Armitage \cite{Ar}. In fact, it is 
this work that we are modeled during our studies.

Since there is only one rank one lattice in $\R$ given by $\Z\hookrightarrow \R$. Hence
$\cM_{\Q,1}[1]$ consists of one point. This means that there is no family but only a 
single Fokker-Planck equation for the torus $\R/\Z$ in classical sense.  

In particular,
\bes
\whx_{\Q,1}(s)=\whz(s)
\ees
where $\whz(s):=\pi^{-s/2}\Ga(s/2)\zeta(s)$ is the complete Riemann zeta function.

Certainly, then the Riemann Hypothesis remains open. So, unlike the situation when
$n\geq 2$, from the beginning, we cannt assume that the zeros of $\whz_{\Q,1}(s)$
has the fowm $\frac{1}{2}+\i\ga\in\frac{1}{2}+i\R$. The idea of Armitage \cite{Ar} and
Berry-Keating \cite{KB} is to use the Fokker-Planck equations in statistical mechanics, 
the Hamiltonians in quantum mechanics, and quantum harmonic oscillators to obtain 
positive definite operators whose eigen values coincide with the zeta zeros, so as to 
materialize an old idea of Bolyai-Hilbert.

To recover Armitage \cite{Ar}, we start with the heat equation for the theta series. Since 
$\cM_{\Q,1}[1]$ consists of one lattice $\Z\subset \R$, for which the dual lattice is itself, 
the metric matrix $H=(1)$ and the Laplacian operator is give  by $\De_{\Z}
=\frac{\partial^2}{\partial x^2}$, the heat equation \eqref{eq.a} is specialized into 
\be
\bc
\frac{1}{4\pi}\frac{\partial^2}{\partial x^2}\theta_\Z(x,t)=\frac{\partial}{\partial t}
\theta_\Z(x,t)\\[0.4em]
\theta_\Z(x,0)=\sum_{n\in\Z}e^{2\pi inx}
\ec
\ee
where the theta function is given by 
$\theta_\Z(x,t)=\sum_{n\in\Z}e^{-\pi i n^2+2\pi i nx}$

Hence if we set $t=\exp(T)$ and $X=\exp(T/2)\,X$, 
\be
\De_{\Z,X}=\frac{\partial^2}{\partial X^2},\quad \lan X,\frac{\partial}{\partial X}\ran=X\frac{\partial}{\partial X}\qan\Theta_Z(X,T)=\sum_{n\in \Z}e^{-\pi e^Tn^2+2\pi i ne^{T/2}X}.
\ee
Hence, the corresponding heat equation \eqref{e1} is specialized as the differential equation on the dual torus 
$\T_{\Z}\simeq\R^n\big/\left(e^{-T/2}\Z\right)$ given by
\be\label{e--1}
\bc
\dis{\left(\frac{\partial }{\partial T}-\Omega_{\Z,X}
\right)}\Big(\Theta_\Z(X,T)\Big)=0,\\[0.80em]
\hskip 1.0cm\Theta_\Z(X,0)=\theta_\La(x,1).
\ec
\ee
Here, $\Omega_{\La^\vee,X}$ denotes the second order differential operator 
defined by
\be\label{eq--0029}
\Omega_{\La^\vee,X}:=\frac{1}{4\pi}\frac{\partial^2}{\partial X^2}
+\frac{1}{2}X\frac{\partial}{\partial X}.
\ee
Consequently, \eqref{e2} takes the form
\small{\bea\label{e--2}
2\left(\whz(s)-\frac{1}{s(s-1)}\right)
=\left(\left(R\left(-\frac{s}{2},\Omega_{\Z,X}\right)
\!+\!R\left(\!-\frac{1-s}{2},\Omega_{\Z,X}\right)\right)\!
\Big(\Theta_\Z(X,0)-\!1\Big)\right)\Big|_{X=0}.
\eea}
Consequently, if we assume that $\frac{1}{2}+i\ga$ is a Riemann zero of $\whz(s)$ on 
the central line, and set
\bea
&\Phi_{\Z}^{(1)}(X,\ga):=\frac{1}{2}\int_0^\infty
\left(e^{(\frac{1}{4}+\frac{1}{2}i\ga)T}+e^{(\frac{1}{4}-\frac{1}{2}i\ga)T}\right)\,
\Big(\Theta_\Z(X,T)-1\Big)dT,\\
&\ka_{\Z}^{(1)}(X):=2\left(2\Omega_{\Z,X}+\frac{1}{2}\right)
\Big(\Theta_{\Z}(X,0)-1\Big).
\eea
and
\bea
&\Phi_{\Z}(X,\ga):=\Phi_{\Z}^{(1)}(X,\ga)+\frac{1}{1/4+\ga^2}\exp(-\pi |X|^2),\\
&\ka_{\Z}(X):=\ka_{\Z}^{(1)}(X)
+\frac{1}{1/4+\ga^2}\left(\left(2\Omega_{\Z,X}+\frac{1}{2}\right)^2+\ga^2\right)
\exp(-\pi |X|^2),
\eea
Hence we may introduce the special value of fat  zeta function by
\bea\label{e11--2}
\left(\left(2\Omega_{\Z,X}+\frac{1}{2}\right)^2+\ga^2\right)&
\left(\whz\Big(X,\frac{1}{2}+i\ga\Big)-\frac{1}
{1/4+\ga^2}\Big(1-\exp(-\pi |X|^2)\Big)\right)\\
=&\left(\left(2\Omega_{\Z,X}+\frac{1}{2}\right)^2
+\ga^2\right)\Phi_{\Z}(X,\ga)\\
=&\ka_{\Z}(X).
\eea
then Theorem 6 becomes

\begin{thm}[Armitage[\cite{Ar}]\label{th---1} Assume that $\dis{frac{1}{2}+i\ga\in\frac{1}{2}+i\R}$ is a 
zero of  $\whz(s)$. Then the function 
$\Phi_{\Z}(X,\ga)$ is a solution of the following \lq average' differential 
equation
\be\label{e--4}
\left(\left(2\Omega_{\Z,X}+\frac{1}{2}\right)^2
+(\ga)^2\right)\Phi_{\Z}(X,\ga)-\ka_{\Z}(X)=0
\ee
and satisfies the following initial conditions
\begin{enumerate}
\item [(1)] $\dis{\Phi_{\Z}(0,\ga)=0}$,
\item [(2)] $\dis{\frac{\partial}{\partial X}
\Phi_{\Z}(0,\ga)
=\frac{\partial^3}{\partial X^3}\Phi_{\Z}(0,\ga)=0}$;
\item [(3)] $\dis{\lim_{|X|\to\infty}\left|\frac{\partial^j}{\partial X^j}
\Phi_{\Z}(X,\ga)\right|=O\left(\frac{1}{|X|}\right)}$ for $0\leq j\leq 4.$
\end{enumerate}
\end{thm}

Consequently, in terms of perturbations, we have the follows.

Suppose there would be a family of Hilbert spaces $\cH_{\Z,0}$ of functions 
defined on the torus  $\T_{e^{-{T}/{2}}\Z}$ with an associated inner product 
$(\cdot,\cdot)$, containing both $\Phi_{\Z}(X,\ga)$ and $\ka_{\Z}(X)$ for all 
$\ga$. Denote the real and the imaginary part of $\Phi_{\Z}(X,\ga)$ by 
$\a_{\Z,\ga}(X)$ and $\b_{\Z,\ga}(X)$. Then
\be
\Phi_{\Z}(X,\ga)=\a_{\Z,\ga}(X)+i\b_{\Z,\ga}(X)\qqan 
\Phi_{\Z}(X,\bar\ga)=\a_{\Z,\ga}(X)-i\b_{\Z,\ga}(X).
\ee
In the spaces $\cH_{\Z,0}$, chose a family of function $\varphi_\Z(X)$ such that
\be
<\b_{\Z,\ga}(X),\varphi_\Z(X)>=0\qqan <\a_{\Z,\ga}(X),\varphi_\Z(X)>\not=0.
\ee
Hence if we  define a family of projection operators 
\be
\begin{matrix}P_{\Z,\ga}:& \cH_{\Z0}&\to &<\ka_{\Z}(X)>,\\[0.5em]
&f&\mapsto&-\dis{\frac{<f,\varphi_\Z(X)>}{<\a_{\Z,\ga}(X),\varphi_\Z(X)>}}
\cdot \ka_{\Z}(X)\end{matrix}
\ee
then
\be
P_{\Z,\bar\ga}=P_{\Z,\ga}.
\ee
and
\be
\left(\left(2\Omega_{\La^\vee,X}+\frac{n}{2}\right)^2+P_{\La,\ga}
+(n\ga)^2\right)\Phi_{\La^\vee}(X,\ga)=0.
\ee
if  $\dis{\frac{1}{2}+\ga}$ is a zero of \,$\whz(s)$ on the central line. 

All in all, we have the following results on the Riemann zeros and the Fokker-Planck equations.

\begin{thm}[Armitage\cite{Ar}] Assume that $\frac{1}{2}+\ga$ is a zero of non-abelian zeta function
$\whz(s)$ on the central line. Then, we have the following equivalent form 
of the Fokker-Planck equations.

\begin{enumerate}
\item [(1)] ~\\[-3.0em]
\small{\bea\label{e--6}
&\left(\!
\frac{\partial}{\partial Y^2} \!-\! \frac{\partial}{\partial Y} Y
\!+\!2\pi \Big(\!Y^2\frac{\partial}{\partial Y}Y
\!+\!\frac{\partial}{\partial Y}Y^3\! \Big)
\!+\! 4\pi^2 Y^4\! -\!2\pi  Y^2\!+\!\Big(\frac{1}{4}+\ga^2\Big)\!\right)
\wh\Phi_{\Z}(Y,\ga)\\
&\hskip 9.0cm -\wh{\ka_{\Z}(Y)}
=0
\eea}
\item[(2)] ~\\[-3.0em]
\bea\label{e6--2}
&\left(\!\!\left(\!\! Y^2
\frac{\partial^2}{\partial Y^2}\!+\!\Big(4\pi Y^2 \!+\! 2\Big)Y\frac{\partial}{\partial Y}\!+\!\Big(Y^2\!+\!1\Big)^2
\!-\!\Big(2\pi Y^2\!+\!1\Big)\!+\!\left(\frac{1}{4}\!+\!\ga^2\right)\right)\right.
\wh\Phi_{\Z}(Y,\ga)\\
&\hskip 9.0cm
-\wh{\ka_{\Z}(Y)}=0
\eea
\item[(3)]~\\[-3.0em]
\bea\label{e6--3}
&\!\!\left(\!\!\left(
\frac{\partial^2}{\partial Y^2} Y^2 
-2 \frac{\partial}{\partial Y} Y
\Big((1+4\pi )Y^2\Big)\!\!
\right)\right.\\
&\left.+2\pi\!\!\left(\!2\pi Y^4\!\!-\!3 Y^2
\!+(Y^2-Y)\Big)\right)+\left(\frac{1}{4}+\ga^2\right)\!\!\right)\wh\Phi_{\Z}(Y,\ga)
  -
\wh{\ka_{\Z}(Y)}=0
\eea
\end{enumerate}
\end{thm}

\subsection{Rank Two}
Rank two case is more complicated. It is the first case that the integration on the base 
moduli space is needed to obtain a uniformly stable theory.

Recall that, from the classical reduction, rank two lattices are parametrized by the 
fundamental domain $\cD$ of $\SL_2(\Z)$ in the upper hale complex plane $\cH$.
More precisely, we have
\bes
\cD:=\left\{z=x+iy\in \C: \bm x^2+y^2\geq 1&- \frac{1}{2}\leq x< \frac{1}{2}\\ 
\& \ \mathrm{if}\ x^2+y^2=1,&x\leq 0\em
\hskip -0.2em \right\}
\ees
This can be used to give an explicit description of the moduli space $\cM_{\Q,2}[1]$
of semi-stable lattice of rank two and volume one. Indeed, by \cite{W}, there is a natural 
identification
\bes
\bm
\cM_{\Q,2}[1]&\simeq& \cD_1:=\big\{z=x+iy\in\cD:\,y\leq 1\big\}\\[0.80em]
\La=\left(\Z^2,\dis{\frac{1}{y}}\bpm x^2+y^2&x\\ x&1\epm\right)&\mapsto& \tau^{~}_\La=x+iy
\em
\ees
In the sequel, we will  simple use $\cD_1$ for $\cM_{\Q,2}[1]$. 

From the definition, the theta function $\theta_\La(x,t)$ of $\La$ is given by
\be
\theta_\La({\bf x};t)=\theta_\La(x_1,x_2;t)=\sum_{(m,n)\in\Z^2}e^{-\pi 
t\frac{(x^2+y^2)m^2+2xmn+n^2}{y^2}}e^{2\pi i(mx_1+nx_2)}.
\ee
This is nothing but the theta function in terms of $\tau=x+yi$ 
\be
\theta_\tau({\bf x};t)=\theta_\tau(x_1,x_2;t)=\sum_{(m,n)\in\Z^2}e^{-\pi 
t\frac{|m\tau+n|^2}{y^2}}e^{2\pi i(mx_1+nx_2)}.
\ee
Similarly, we have
\be
\theta_{\La^\vee}({\bf x};t)=\sum_{(m,n)\in\Z^2}e^{-\pi 
t\frac{m^2-2xmn+(x^2+y^2)n^2}{y^2}}e^{2\pi i(mx_1+nx_2)}.
\ee
which coincides with  the theta function in terms of $\tau^\vee=-x+yi$ 
\be
\theta_{\tau^\vee}({\bf x};t)=\sum_{(m,n)\in\Z^2}e^{-\pi 
t\frac{|m-n\tau|^2}{y^2}}e^{2\pi i(mx_1+nx_2)}.
\ee

We notice that in fact
\be
\theta_{\La^\vee}({\bf x};t)=\theta_{\La}({\bf x};t)\qqan 
\theta_{\tau_\La}({\bf x};t)=\theta_\tau({\bf x};t)
\ee
which can be obtained by  simple changes of  $\pm m$ and $\mp n$  within 
the pairs $(m,n)\in\Z^2$. This is not just an coincidence. Indeed,  since the dual lattice 
$\La^\vee$ of $\La$ and its corresponding point in $\cH$ are 
\bes
\La^\vee=\left(\Z^2,\dis{\frac{1}{y}}\bpm 1&-x\\ -x&x^2+y^2\epm\right)
\mapsto\tau_{\La^\vee}
=-\frac{x}{x^2+y^2}+\frac{y}{x^2+y^2}i=:\tilde x+i\tilde y
\ees
which is not a point of $\cD$,
the associated Laplacian operator on differential forms of $\T_{\La^\vee}$ becomes
\bes
\De_{\La^\vee,\bf x}=\frac{1}{y}\left(\frac{\partial^2}{\partial x_1^2} -2x
\frac{\partial^2}{\partial x_1\partial x_2}+(x^2+y^2)\frac{\partial^2}{\partial x_2^2}\right),
\ees
or better, in terms of $\tau_{\La^\vee}$
\bea
\De_{\tau^\vee}=&\frac{1}{\tilde y}\left(\frac{\partial^2}{\partial x_1^2} -2\tilde x
\frac{\partial^2}{\partial x_1\partial x_2}+({\tilde x}^2+{\tilde y}^2)\frac{\partial^2}{\partial x_2^2}\right)\\
=&\frac{x^2+y^2}{y}\left(\frac{\partial^2}{\partial x_1^2} +2\frac{ x}{x^2+y^2}
\frac{\partial^2}{\partial x_1\partial x_2}+\left(\frac{(-x)^2}{(x^2+y^2)^2}+\frac{y^2}{(x^2+y^2)^2}\right)\frac{\partial^2}{\partial x_2^2}\right)\\
=&\frac{1}{y}\left(({x}^2+{y}^2)\frac{\partial^2}{\partial x_1^2} +2x
\frac{\partial^2}{\partial x_1\partial x_2}+\frac{\partial^2}{\partial x_2^2}\right)=\De_\tau
\eea
In addition, note that $\dis{\tau_{\La^\vee}=\frac{-x+iy}{x^2+y^2}=\frac{-\bar\tau_\La}
{\tau_\La\bar\tau_\La}=-\frac{1}{\tau_\La}}$ which is the fractional transform image of 
$\tau_\La$ under the element $\bpm 0&1\\ -1&0\epm$. Therefore, $\tau_\La$ and 
$\tau_{\La^\vee}$ are $\SL_2(\Z)$-equivalent. This implies that  both the lattices $\La$ 
and $\La^\vee$  correspond to the same moduli point $\tau=\tau_\La\in\cD_1$. 
Consequently, $\La\simeq\La^\vee$, or equivalently, the matrices 
$H_\La=\dis{\frac{1}{y}}\bpm x^2+y^2&x\\ x&1\epm$ and 
$H_{\La^\vee}=\dis{\frac{1}{y}}\bpm 1&-x\\ -x&x^2+y^2\epm$ are $\SO(2)$-equivalent.
For this reason, we will only use $\La$ instead of both $\La$ and its dual  $\La^\vee$.

Obviously, the heat equation form the theta function  $\theta_\La(x_1,x_2;t)$ 
becomes
\be
\bc
\dis{\frac{1}{4\pi}\Delta_{\La,x} \Big(\theta_\La(x,t)\Big)
=\frac{\partial }{\partial t}\Big(\theta_\La(x,t)\Big)}\\[0.5em]
\theta_\La(x,0)=\sum_{(m,n)\in\Z^2}e^{2\pi i(mx_1+nx_2)}
\ec\ee
on the torus $\T_{\La}$. In terms of $\tau_\La\in\cD_1$, we have
\be
\bc
\dis{\frac{1}{4\pi}\Delta_{\tau,x} \Big(\theta_\tau(x,t)\Big)
=\frac{\partial }{\partial t}\Big(\theta_\tau(x,t)\Big)}\\[0.5em]
\theta_\tau(x,0)=\sum_{(m,n)\in\Z^2}e^{2\pi i(mx_1+nx_2)}
\ec\ee
Here, we have used the fact that
\bes
\De_{\La,x}=\De_{\tau,x}=\frac{1}{y}\left((x^2+y^2)\frac{\partial^2}{\partial x_1^2} +2x
\frac{\partial^2}{\partial x_1\partial x_2}+\frac{\partial^2}{\partial x_2^2}\right).
\ees

To obtain the Fokker-Planck equation, we make a change of variables $t=:\exp(T)$, 
$(x_1,x_2)=:\exp(T/2)(X_1,X_2)=:\exp(T/2)X$. Accordingly, we define
\bea
\Delta_{\La,X}:=&\Delta_{\tau,X}:=\frac{1}{y}\left((x^2+y^2)\frac{\partial^2}{\partial X_1^2} +2x
\frac{\partial^2}{\partial X_1\partial X_2}+\frac{\partial^2}{\partial X_2^2}\right),\\
 \Big\lan X,\frac{\partial}{\partial X}\Big\ran
:=&X_1\frac{\partial }{\partial X_1}+X_2\frac{\partial }{\partial X_2}.
\eea
and
\bea
\Theta_\La(X,T):=&\Theta_\tau(X,T):=\theta_\La\left(X\exp(T/2),\exp T\right)\\
=&\sum_{(m,n)\in\Z^2}e^{-\pi e^{T}\frac{|m\tau+n|^2}{y^2}}e^{2\pi i(mX_1+nX_2) 
e^{T/2}}.
\eea
Then, by \eqref{e1},  on the dual torus $\T_{\La}\simeq\R^2\big/\left(e^{-T/2}\La\right)$, 
we have the following heat equation for the Theta function
\be\label{1e}
\bc
\dis{\left(\frac{\partial }{\partial T}-\Omega_{\tau,X}
\right)}\Big(\Theta_\tau(X,T)\Big)=0,\\[0.50em]
\hskip 2.750cm \Theta_\tau(X,0)=\theta_\tau(x,1).
\ec
\ee
Here, $\Omega_{\tau,X}$ is a second order differential operator defined by
\be
\Omega_{\tau,X}:=\Omega_{\La,X}:=\frac{1}{4\pi}\Delta_{\tau,X}
+\frac{1}{2}\Big\lan X,\frac{\partial}{\partial X}\Big\ran.
\ee
Consequently, if we denote by $R(\kappa,\Omega_{\tau,X})$ the resolvent 
associated to $(-\kappa-\Omega_{\tau,X})^{-1}$, then for the rank $2$ non-abelian zeta 
function $\whz_{\Q,2}(s)$,
\small{\bea\label{2e}
&\left(\whz_{\Q,2}(s)-\frac{1}{s(s-1)}\vol(\cD_1)\right)\\
=&\int_{\cD_1}\!\!\!\left(\left(R\left(-s,\Omega_{\tau,X}\right)
\!+\!R\left(\!-(1-s),\Omega_{\tau,X}\right)\right)\!
\Big(\Theta_\tau(X,0)-\!1\Big)\right)\Big|_{X=0}\!\frac{dx\wedge dy}{y^2}.
\eea}

Now, let $\frac{1}{2}+i\ga$ be a zero of $\whz_{Q,2}(s)$. We introduce the auxiliary 
functions $\Phi_{\tau,X}^{(1)}(X,\ga)=\Phi_{\La}^{(1)}(X,\ga)$ and 
$\ka_{\tau}^{(1)}(X)=\ka_{\La}^{(1)}(X)$  by
\bea
\Phi_{\tau}^{(1)}(X,\ga):=&\int_0^\infty
\left(e^{(\frac{1}{2}+i\ga)T}+e^{(\frac{1}{2}-i\ga)T}\right)\,
\Big(\Theta_\tau(X,T)-1\Big)dT,\\
\ka_{\tau}^{(1)}(X):=&4\left(2\Omega_{\tau,X}+1\right)
\Big(\Theta_{\tau}(X,0)-1\Big).
\eea
and the fat zeta function $\whz_{\Q,2}\Big(X,\frac{1}{2}+i\ga\Big)$ by
\bea\label{2e11}
&\left(\left(2\Omega_{\tau,X}+\right)^2+4\ga^2\right)
\left(\whz_{\Q,2}\Big(X,\frac{1}{2}+i\ga\Big)-\frac{\vol(\cD_1)}{1/4+\ga^2}\Big(1-\exp(-\pi |X|^2)\Big)\right)\\
:=&\int_{\cD_1}\left(\left(\left(2\Omega_{\tau,X}+1\right)^2
+4\ga^2\right)\Phi_{\tau}(X,\ga)\right)\frac{dx\wedge dy}{y^2}\\
=&\int_{\cD_1}\ka_{\tau}(X)\frac{dx\wedge dy}{y^2}.
\eea
Hence, by Theorem\,\ref{th1},  we have the following

\begin{thm}\label{1th} Assume that $\frac{1}{2}+i \ga$ is a zero of non-abelian zeta 
function $\whz_{\Q,2}(s)$ on the central line.Then
$\Phi_{\tau}(X,\ga)$ is a solution of the following average differential equation
\be\label{44e}
\int_{\cD_1}\left(\left(\left(2\Omega_{\tau,X}+1\right)^2+4\ga^2\right)\Phi_{\tau}(X,\ga)-\ka_{\tau}(X)\right)\,\frac{dx\wedge dy}{y^2}=0
\ee
and satisfies the following initial properties
\begin{enumerate}
\item [(1)] $\dis{\int_{\cD_1}\Phi_{\tau}(0,\ga)\,\frac{dx\wedge dy}{y^2}=0}$,
\item [(2)] $\dis{\int_{\cD_1}\frac{\partial}{\partial X_i}
\Phi_{\tau}(0,\ga)\,\frac{dx\wedge dy}{y^2}
=\int_{\cD_1}\frac{\partial^3}{\partial X_i^3}\Phi_{\tau}(0,\ga)\,\frac{dx\wedge dy}{y^2}=0}$;
\item [(3)] $\dis{\lim_{|X|\to\infty}\left|\int_{\cD_1}\frac{\partial^j}{\partial X_i^j}
\Phi_{\tau}(X,\ga)\,\frac{dx\wedge dy}{y^2}\right|=O\left(\frac{1}{|X|}\right)}$ for $0\leq j\leq 4.$
\end{enumerate}
\end{thm}

Consequently, using the projection operator $P_{\La,\ga}$, we have
\be
\int_{\cD_1}
\left(\left(\left(2\Omega_{\tau,X}+1\right)^2+P_{\La,\ga}+4\ga^2\right)
\Phi_{\tau}(X,\ga)\right)\,\frac{dx\wedge dy}{y^2}=0.
\ee
provided that $\dis{\frac{1}{2}+\ga i}$ be a zero of \,$\whz_{\Q,2}(s)$, which is known on the central line. 

All these then lead to the following

\begin{thm} Assume that $\frac{1}{2}+i \ga$ is a zero of non-abelian zeta function
$\whz_{\Q,2}(s)$ on the central line. Then, we have the following equivalent form 
of the average of Fokker-Planck equations over $\cM_{\Q,2}[1]$.

\begin{enumerate}
\item [(1)] The  Fokker-Planck equation over the universal family $\pi:\cT\to \cD_1$  of tori is given by
\bea\label{6e}
\int_{\cD_1}
&\left(\left(
\frac{\partial}{\partial Y_1}
Y_1\frac{\partial}{\partial Y_1} Y_1+2\frac{\partial}{\partial Y_1}
Y_1\frac{\partial}{\partial Y_2} Y_2+\frac{\partial}{\partial Y_2}
Y_2\frac{\partial}{\partial Y_2} Y_2-2 \frac{\partial}{\partial Y_1} Y_1-2\frac{\partial}{\partial Y_2} Y_2\right.\right.\\
&+\frac{2\pi}{y} \Big((x^2+y^2)
\Big(Y_1^2\frac{\partial}{\partial Y_1}Y_1 +\frac{\partial}{\partial Y_1}
Y_1^3\Big) +(x^2+y^2)
\Big(Y_1^2\frac{\partial}{\partial Y_2}Y_2 +\frac{\partial}{\partial Y_2}
Y_1^2 Y_2\Big)\\
&\qquad\quad+2x
\Big(Y_1 Y_2\frac{\partial}{\partial Y_1}Y_1 +\frac{\partial}{\partial Y_1}
Y_1^2 Y_2 \Big) +
\Big(Y_1Y_2\frac{\partial}{\partial Y_2}Y_2 +\frac{\partial}{\partial Y_2}
Y_1 Y_2^2\Big)\Big)\\
&\qquad\quad+
Y_2^2\Big(\frac{\partial}{\partial Y_1}Y_1 +\frac{\partial}{\partial Y_1} Y_1\Big) +
\Big(Y_2^2\frac{\partial}{\partial Y_2}Y_2 +\frac{\partial}{\partial Y_2}
Y_2^3\Big)\Big)\\
&+\frac{4\pi^2}{y^2}\Big((x^2+y^2)Y_1^2+2xY_1Y_2+Y_2^2\Big)^2 
-\frac{4\pi}{y}   \Big((x^2+y^2)Y_1^2+2xY_1Y_2+Y_2^2\Big)\\
&\hskip 3.70cm \left.+4\left(\frac{1}{4}+\ga^2\right)\Big)
\wh{\Phi_{\tau}(Y,\ga)} -\wh{\ka_{\tau}(Y)}
\right)\frac{dx\wedge dy}{y^2}=0
\eea

\item[(2)] The forward Fokker-Planck equation over the universal family $\pi:\cT\to \cD_1$  of tori is given by
\bea\label{2e6}
\int_{\cD_1}&\!\!\left(\!\!\left(
\Big(Y_1^2 \frac{\partial^2}{\partial Y_1^2}+2Y_1 Y_2
\frac{\partial^2}{\partial Y_1\partial Y_2}+Y_2^2
\frac{\partial^2}{\partial Y_2^2}\Big)\right.\right.\\
+&\left(\frac{4\pi}{y}\Big((x^2+y^2)Y_1^2+2xY_1Y_2+ Y_2^2\Big) + 3\right)
\left(Y_1\frac{\partial}{\partial Y_1}+Y_2\frac{\partial}{\partial Y_2}\right)\\
+&\Big(\Big(\frac{4\pi}{y}\Big((x^2+y^2)Y_1^2+2xY_1Y_2+ Y_2^2\Big) +1\Big)^2
-\Big(\frac{2\pi}{y}\Big((x^2+y^2)Y_1^2+Y_2^2\Big)+1\Big)\\
&\hskip 3.70cm  \left.\left.+4\left(\frac{1}{4}+\ga^2\right)\right)
\wh{\Phi_{\tau}(Y,\ga)} - \wh{\ka_{\tau}(Y)}\right)\frac{dx\wedge dy}{y^2}=0\\
\eea
\item[(3)] The backward Fokker-Planck equation over the universal family $\pi:\cT\to \cD_1$ of tori is given by
\bea\label{3e6}
\int_{\cD_1}&
\left(\left(
\Big(
\frac{\partial^2}{\partial Y_1^2} Y_1^2 +2\frac{\partial^2}{\partial Y_1\partial Y_2} Y_1 Y_2+\frac{\partial^2}{\partial Y_2} Y_2^2
\Big)\right.\right.\\ 
&\qquad\qquad\left. 
-3 \Big(
\frac{\partial}{\partial Y_1} Y_1+\frac{\partial}{\partial Y_2} Y_2
\Big)
\Big(
1+\frac{4\pi}{y}
\Big(
(x^2+y^2)Y_1^2+2xY_1Y_2+Y_2^2
\Big)
\Big)\right)\\
&+2\pi
\Big(
\frac{2\pi}{y^2}
\Big(
(x^2+y^2)Y_1^2+2xY_1Y_2+Y_2^2
\Big)^2\!\!-\!\frac{4}{y}\!
\Big(
(x^2+y^2)Y_1^2+2xY_1Y_2+Y_2^2
\Big)
\\
&\qquad\qquad
\!+\Big(
(Y_1^2+Y_2^2)-(Y_1+Y_2)
\Big)
\Big)\\
&\hskip 3.70cm \left.\left.
+4
\left(
\frac{1}{4}+\ga^2
\right)
\right) \wh{\Phi_{\tau}(Y,
\ga)}-
\wh{\ka_{\tau}(Y)}
\right)
\frac{dx\wedge dy}{y^2}=0
\eea\end{enumerate}
\end{thm}

Apparently very complicated,   it is clear that these different equations are of the 
second order and elliptic on the fiber direction of the fibration $\pi:\cT\to \cD_1$, which 
becomes stable under an average on the base $\cD_1$. In this sense, $1+4\ga^2$ 
appears as the eigenvalue for the eigenfunction $\\wh{\Phi_\tau(Y,\ga)}$ of the 
corresponding second order elliptic operators.

\section{Moduli Spaces of Abelian Varieties and Curves}

There are natural generalizations of non-abelian zeta functions to the zeta functions of 
the special linear groups $\SL_n$ and their maximal parabolic subgroups 
$P_{n_1,n_2,\ldots,n_k}$ determined by ordered partitions $n=n_1+n_2+n_k$, and 
more generally, to the zeta functions of pairs consisting of a split reductive group and 
its maximal parabolic subgroup. One of the main purpose of \cite{W} is to construct 
these zeta functions and study their basic properties. Best of all, we establish there a 
weak version for the Riemann hypothesis for all these zeta functions when working 
over the field of rationals, claiming that all but finitely many zeros of these zeta 
functions lie on the central line $\dis{\Re(s)=\frac{1}{2}}$, provided that the rank of all 
its simple factors of $G$ is at least one of course (hence  the original 
Riemann hypothesis for the Riemann zeta function is not included). Therefore, the  
discussions above have their companions the lattices $\La$ replacing by the lattices in
$\Lie(G)_\R$ associated to what we call compatible arithmetical $G$-torsors 
and the space $\cM_{\Q,n}$ by the moduli spaces of 
semi-stable arithmetic $G$-torsors of slope zero

In addition, probably more challenging, we can work over the moduli spaces 
$\cA_{\Q,n}$ of abelian varieties whose structural lattices are semi-stable. This new 
moduli space is a compact subspaces of the moduli spaces of polarized abelian 
varieties of dimension $n$. In this acse, certainly, the reductive group $G$ should be 
taken to be the symplectic group $\Sp_{2n}$. Similar structure as above should exist. 
What are the implications of these new structures

There are too many questions to be asked. For one,  we may yet further construct  a 
new type of zeta functions by working over the moduli spaces $\cC_{\Q,g}$ of regular 
curves  of genus $g$ for which the lattices of their Jacobians are assumed to be 
semi-stable. The moduli spaces $\cC_{\Q,g}$ may be viewed as a subspaces in the 
$\cA_{\Q,n}$, via Jacobean embeddings, or better, subspaces of the moduli spaces of 
curves $\cM_p$. With  the kdV equation characterizations for the theta functions of 
curves, we expect that there is a much more refined structures here as well.

\appendix
\section{Fokker-Planck Equation}
\subsection{One Variable}

The single variable {\it Fokker-Planck equation} with time-independent drift  and 
diffusion coefficients $D^{(1)}(x)$ and $D^{(2)}(x)$ is given by
\be\label{fp1}
\frac{\partial}{\partial t}W(x,t)=L_{\mathrm{FP}}W(x,t)
\ee
where the Fokker-Planck (differential) operator is defined by
\be
L_{\mathrm{FP}}=-\frac{\partial}{\partial x}D^{(1)}(x)
+\frac{\partial^2}{\partial x^2}D^{(2)}(x).
\ee
In many cases, the Fokker-Planck equation \eqref{fp1} satisfied by the distribution 
function $W(x,t)$ characterizes the time-dependent distribution function.

\begin{ex}\normalfont When  a small particle of mass $m$ of the velocity $v(t)$ is 
immersed in a fluid, because of the collisions between the molecules of the fluid and 
the particle, the fraction force arises.  Accordingly, the momentum of the particle is 
transferred to the molecules of the fluid and the velocity of the particle gradually 
decreases to zero. To describe this phenomenon, we first assume that the mass of the 
particle is large enough so that its velocity due to thermal fluctuations is negligible.
Then the resulting fraction force is, accordingly to Stokes's raw, given by
\be
F_c(t)=-\a v(t)
\ee
where $\a$ is a constant depending on the material. Since this fraction force can be 
written in terms of acceleration $\dot v(t)$ by
\be\label{1.2}
F_c(t)=m\dot v(t).
\ee
All these then leads to the equation of motions
\be\label{1.3}
m\dot v(t)+\a v(t)=0.
\ee
Therefore, the motion is completely determined by the relation
\be
v(t)=v(0)\,e^{-\frac{\a}{m}t}.
\ee

However, when the mass becomes very small, the thermal velocity $v_{\mathrm{th}}$  
is observed. Since, by the equipartition law, the mean energy of the particle is given by
\be
\frac{1}{m}\lan v^2\ran=\frac{1}{2}\ga T
\ee
where $\ga$ is the Boltzmann constant and $T$ is the temperature.\footnote{For the 
meaning of $\lan\ \ran$, please refer \eqref{eqoo76}.} Hence, the thermal velocity is 
calculated by
\be
v_{\mathrm{th}}=\sqrt{\lan v^2\ran}=\sqrt{\frac{\ga T}{m}}.
\ee
In particular, the smaller the mass $m$ is, the more  the thermal velocity 
$v_{\mathrm{th}}$ becomes observable. As a result, for small mass particles, their 
velocities cannot be characterized by \eqref{1.3}. 

Nevertherless, no matter how small, when the mass of the particle is still bigger than
the mass of the molecules,  \eqref{1.3} can be modified to  correct thermal energy
by adding a fluctuating force $F_{\mathrm{fl}}(t)$, the so-called {\it Langevin force}. 
That is to say, instead of \eqref{1.2}, we get
\be
-m\dot v(t)=F(t)=F_c(t)+F_{\mathrm{fl}}(t)=-\a v(t)+F_{\mathrm{fl}}(t)
\ee
 
In particular, if we denote the the Langevin force by 
$\dis{\Ga(t):=\frac{1}{m}\,F_{\mathrm{fl}}(t)}$, the relation \eqref{1.3} is modified to
\be
\dot v(t)+\frac{\a}{m}\, v(t)=\Ga(t).
\ee

Since the force $F_{\mathrm{fl}}(t)$ is a stochastic or a random force, their properties 
are only given in the average. In terms of the Langevin force $\Ga(t)$, since the motion 
of the average velocity $\lan v(t)\ran$ should be given by \eqref{1.2}, we may assume 
that both the average of $\Ga(t)$ and the average of the correlation of two Langevin 
forces for time differences $t'-t$ which are larger than the duration time $\tau_0$ of a 
collision are zero, i.e.
\be
\lan\Ga(t)\ran=0\qan \lan\Ga(t)\Ga(t')\ran=0\quad \forall |t-t_0|\geq \tau_0.
\ee
In physics, the second relation then finally leads to the relation that
\be
\lan\Ga(t)\Ga(t')\ran=2\frac{\a \ga T}{m^2}\,\delta(t-t')
\ee
where $\delta$ denotes the Dirac $\delta$-distribution.

Moreover, since $\Ga(t)$, varying from system to system in the ensemble, is a stochastic quantity, and so is the velocity. Its distribution function, or the same, the probability density, $W(v, t)$ is known to satisfies the one-variable Fokker-Planck equation
\be\label{1.13}
\frac{\partial W}{\partial t}=\frac{\a}{m}\frac{\partial (v\,W)}{\partial v}+
\frac{\a \ga T}{m^2}\frac{\partial^2 W}{\partial v^2}.
\ee 
Here, being a probability density, we have, 
for any function $g(v)$ of $v$, 
\be\label{eq0076}
\lan g(v(t))\ran=\int_{-\infty}^\infty g(v) W(v,t)\,dv.
\ee
\end{ex}

The Fokker-Planck equation \eqref{1.13} is one of the simplest forms, with constant
coefficients $\dis{\frac{\a}{m}}$ and  $\dis{\frac{\a \ga T}{m^2}}$ of the first order and 
second order partial differentials, respectively. General Fokker-Planck equationa are 
taken the form 
\be\label{4.14f}
\frac{\partial W(x,t)}{\partial t}=
\left(-
\frac{\partial}{\partial x}D^{(1)}(x)+
\frac{\partial^2 }{\partial x^2}D^{(2)}(x)
\right) W(x,t).
\ee
For our own conference, we call \eqref{4.14f} a forward Fokker-Planck equation.
Similarly, by a backward Fokker-Planck equation, we mean that of the form
\be\label{4.14b}
\frac{\partial W(x,t)}{\partial t}=
\left(-D^{(1)}(x)
\frac{\partial}{\partial x}+D^{(2)}(x)
\frac{\partial^2 }{\partial x^2}
\right) W(x,t).
\ee

To see how these equations arise, let us consider a general Langevin equation on one 
stochastic variable $\xi$
\be\label{3.67}
\dot\xi=h(\xi,t)+g(\xi,t)\,\Ga(t)
\ee
such that the  Langevin force (with multiplicative noise) is bounded by the conditions 
that
\be
\lan\Ga(t)\ran=0\qqan\lan\Ga(t)\Ga(t')\ran=2\delta(t-t').
\ee
Here multiplicative refers to the fact that $g(\xi,t)$ is not a constant in which case, we 
get an additive noise. But such a noise can be easily treated, since the difference 
between the multiplicative and additive noises are not that significant because, a 
simple change of variables would imply
\be
\dot \eta=\frac{\dot\xi}{g}=\frac{h}{g}+\Ga(t).
\ee

Assume \eqref{3.67}. If we introduce the Kramer-Moyer expansion 
coefficients $D^{(n)}(x,t)$ by
\be\label{eq670}
D^{(n)}(x,t):=\frac{1}{n!}\lim_{\tau\to 0}\frac{\lan [\xi(t+\tau)-x]^n\ran}{\tau}\Big|_{\xi(t)=x},
\ee
then
\be
D^{(n)}(x,t)=
\bc
\dis{h(x,t)+\frac{\partial g(x,t)}{\partial x}g(x,t)}&n=1,\\[0.50em]
g^2(x,t)&n=2\\[0.30em]
0&n\geq 3
\ec
\ee
In physics term, $h(x,t)$ is called the deterministic drift and the so-called 
noise-induced drift contained in $D^{(1)}$ is defined by
\be
D^{(1)}_{\mathrm{ind}}:=\frac{\partial g(x,t)}{\partial x}g(x,t)
=\frac{1}{2}\frac{\partial}{\partial x}D^{(2)}(x,t).
\ee

\subsection{Several Variables}

Let $\xi=(\xi_1,\ldots,\xi_n)$ be $N$ stochastic variables satisfying the following syatem 
of Langevin equations
\be\label{3.067}
\bc
\dot \xi_1=h_1(\xi)+\sum_{\b=1}^ng_{1j}(\xi,t)\Ga_j(t)\\[0.30em]
\quad\dotfill\quad\\[0.30em]
\dot \xi_n=h_n(\xi)+\sum_{\b=1}^ng_{nj}(\xi,t)\Ga_j(t)
\ec
\ee
subject the following constrains
\be
\lan\Ga_i(t)\ran=0\qqan \lan\Ga_i(t)\Ga_j(t')\ran=2\delta_{ij}\,\delta(t-t')
\ee
As in \eqref{eq670}, we introduce the Kramer-Moyal coefficients
\be
D_{i_1\ldots i_\nu}(x,t):=D_{i_1\ldots i_\nu}^{(\nu)}(x,t):=\frac{1}{\nu!}
\lim_{\tau\to 0}\frac{1}{\tau}\lan[\xi_{i_1}(t+\tau)-x_{i_1}]\ldots[\xi_{i_\nu}(t+\tau)
-x_{i_\nu}]\ran.
\ee
In particular, the drift and diffusion coefficients are hence given by
\be
\bc
D_i(x,t)=&D_i^{(1)}(x,t):=\lim_{\tau\to 0}\frac{\lan\xi_i(t+\tau)-x_i\ran}{\tau}\Big|_{\xi_k(t)=x_k}\qquad k=1,2,\ldots, n\\[0.50em]
D_{ij}(x,t)=&D_{ij}^{(2)}(x,t):=\frac{1}{2}\lim_{\tau\to 0}\frac{\lan[\xi_i(t+\tau)-x_i][\xi_j(t+\tau)-x_j]\ran}{\tau}\Big|_{\xi_k(t)=x_k} k=1,2,\ldots, n\\
\ec
\ee
It is not too difficult then to conclude that
\be
\bc
D_i(x,t)&=\dis{h_i(x,t)+\sum_{k,\b=1}^n g_{ki}(x,t)\frac{\partial}{\partial x_k}g_{ij}(x,t)}\\[0.60em]
D_{ij}(x,t)&=\dis{\sum_{k=1}^n g_{ik}(x,t)g_{jk}(x,t)}\\[0.30em]
D_{i_1\ldots i_\nu}(x,t)&=0\hskip 2.0cm (\nu\geq 3)
\ec
\ee
Similar as in one variable case, in fact, the drift and diffusion coefficients determine the Langevin forces, deterministic drifts $h_i(x,t)$ and noise-induced drifts $g_{ij}(x,t)$.
Indeed, if we set $D$ be  matrix $D=(D_{ij})$, then
\be
\bc
g_{ij}=(D^{\frac{1}{2}})_{ij},\\[0.50em]
\dis{h_i=D_i-\sum_{k,\b=1}^n(D^{\frac{1}{2}})_{kj}\frac{\partial}{\partial x_k}(D^{\frac{1}{2}})_{ij}}
\ec
\ee
Here, we have use the fact that $D=(D_{ij})$ is a the symmetric positive definite matrix, 
and hence its square root makes sense (by the taking positive of the square of its 
eigenvalues).

Since the Langevin equation \eqref{3.67} with $\delta$-correlated Langevin forces is a Markov process, namely, its conditional probability at time $t_n$ depends only on the variable $\xi(t_{n-1})=x_{n-1}$ at the next earlier time,  i.e.
\be\label{2.73}
P(x_n,t_n|x_{n-1},t_{n-1};\ldots; x_1,t_1)=P(x_n,t_n|x_{n-1},t_{n-1})W_{n-1}(x_{n-1},t_{n-1};\ldots; x_1,t_1),
\ee  
we get the following Fokker-Planck equation, or the same, the forward Kolmogorov equation
\be
\left(\frac{\partial}{\partial t}-L_{\mathrm{FP}}(x,t)\right)W(x,t)=0
\ee
where
\be
L_{\mathrm{FP}}(x,t)
=-\sum_{\a=1}^N\frac{\partial}{\partial x_i}D_i(x,t)+\sum_{\a,\b=1}^n\frac{\partial^2}
{\partial x_i\partial x_j}D_{ij}(x,t).
\ee

In general, te Langevin equation \eqref{3.67} does not gives rise to a Markov process, 
it is well known that by introducing  new random variables, non-Markovian process can
be deduced from a Markov one. In this sense, the Fokker-Planck equation is really a 
general law. In fact these equations play very important roles in various branches of 
disciplines.

The point we want to make in this paper is that while the above discussions work well 
for a fixed ensemble for which the systems lie on, when there is a family of ensembles, 
the Langevin forces does not act in a simply way as if the ensembles involved would be 
totally independent. Contrary to this, the family of Langevin forces  interact with each 
other within the family so that only after 
taking an average on the base space over which the ensembles form a family. Indeed,
as what we observe in the main text, for the Fokker-Planck equation to stand on, there 
should be an average  on the base space in force. In other words, there is no a single 
Langevin equation which dominates each ensemble in a family, but a global type of  
relations on taking the averages of family Langevin forces over the parametrized 
spaces. We call such an equation a global average force equations. This is the 
essence of our current work.

\eject

~\vskip 11.0cm

Lin WENG

Institute for Fundamental Research

$L$-Academy

and

Faculty of Mathematics 

Kyushu University 

Fukuoka, 819-0395

JAPAN

E-Mail: weng@math.kyushu-u.ac.jp


\begin{thebibliography}{80}


\bibitem{Ar} J.V. Armitage, The Riemann hypothesis and the Hamiltonian of a quantum mechanical system, in \lq\lq {\it Number Theory and Dynamic Systems}", London Math. Soc. Lecture Series 134, Cambridge Univ. Press 1989, pp. 153-172

\bibitem{BK} M.V.Berry and J.P.Keating, The Riemann Zeros and Eigenvalue Asymptotics, SIAM Review, 41 (2): 236-266

\bibitem{K} O. Kallenberg, {\it Foundations of Modern Probability}.  2ed edition. Probability and its Applications. Springer-Verlag  2002. xx+638 pp.
\bibitem{R} H. Risken, {\it The Fokker-Planck Equation. Methods of solution and applications}. 2ed edition. Springer Series in Synergetics 18. Springer-Verlag 1989. xiv+472 pp
\bibitem{W} L. Weng, {\it Zeta Functions of Reductive Groups and Their Zeros},  World Scientific 2018, xxviii+528 pp.

\end{thebibliography}
\end{document}